\documentclass[twocolumn]{aastex62}
\usepackage{amsmath}
\graphicspath{{./}{figures/}}

\received{10, 2018}
\accepted{11, 2018}
\submitjournal{AJ}

\shorttitle{Probabilistic Random Forest}
\shortauthors{Reis, Baron \& Shahaf}

\begin{document}

\title{Probabilistic Random Forest: A machine learning algorithm for noisy datasets}

\correspondingauthor{IR and DB }
\email{itamarreis@mail.tau.ac.il, dalyabaron@gmail.com}

\author{Itamar Reis}
\affiliation{School of Physics and Astronomy \\
Tel-Aviv University \\ 
Tel Aviv 69978, Israel.}
\nocollaboration

\author{Dalya Baron}
\affiliation{School of Physics and Astronomy \\
Tel-Aviv University \\ 
Tel Aviv 69978, Israel.}
\nocollaboration

\author{Sahar Shahaf}
\affiliation{School of Physics and Astronomy \\
Tel-Aviv University \\ 
Tel Aviv 69978, Israel.}
\nocollaboration



\begin{abstract}

Machine learning (ML) algorithms become increasingly important in the analysis of astronomical data. However, since most ML algorithms are not designed to take data uncertainties into account, ML based studies are mostly restricted to data with high signal-to-noise ratio. Astronomical datasets of such high-quality are uncommon. In this work we modify the long-established Random Forest (RF) algorithm to take into account uncertainties in the measurements (i.e., features) as well as in the assigned classes (i.e., labels). To do so, the Probabilistic Random Forest (PRF) algorithm treats the features and labels as probability distribution functions, rather than deterministic quantities. We perform a variety of experiments where we inject different types of noise to a dataset, and compare the accuracy of the PRF to that of RF. The PRF outperforms RF in all cases, with a moderate increase in running time. We find an improvement in classification accuracy of up to 10\% in the case of noisy features, and up to 30\% in the case of noisy labels. The PRF accuracy decreased by less then 5\% for a dataset with as many as 45\% misclassified objects, compared to a clean dataset. Apart from improving the prediction accuracy in noisy datasets, the PRF naturally copes with missing values in the data, and outperforms RF when applied to a dataset with different noise characteristics in the training and test sets, suggesting that it can be used for Transfer Learning. 

\end{abstract}



\section{Introduction} \label{sec:intro}

Machine learning (ML) algorithms have become important tools for analyzing astronomical datasets and are used for a wide variety of tasks. Machine learning (ML) algorithms were found to be useful in cases where, due to the increasing size and complexity of the data, traditional methods such as visual inspection of the data or model fitting are becoming impractical. In astronomy, ML is commonly used in a supervised setting \citep[e.g,][]{banerji10, masci14, kim15, moller16, parks18, zucker18}. A supervised ML algorithm is used to learn a relationship from a set of examples. The learned relationship can then be applied for prediction on unseen data. 

In ML terminology, the dataset is composed of objects, each object having features, and a target variable. A relationship is learned between the features and the target variable. When the target variable is discrete the ML task is referred to as classification \citep[e.g,][]{richards11, bloom12, brink13, djorgovski14, mahabal17, castro18}, and when the target variable is continuous the task is referred to as regression \citep[e.g,][]{dlsanto18, dlsanto18b}. In astronomy the objects are usually physical entities like galaxies or stars, and the features are usually measurements like spectra or light curves, or higher level quantities derived from the actual measurements (e.g., variability periods, or emission line width). In a classification task, the target variable can be a label such as the galaxy type. In a regression task, it can be a physical property such as the stellar mass. 

Supervised ML is required in cases where our best physical or phenomenological model of the relationship is not general enough to provide a good description of the relationship that we seek. However, it is used even in cases where an adequate model exists, merely for being easier to implement and faster to run. This is due to the availability of many powerful off-the-shelf ML algorithms that are open-source and publicly-available for general use cases (see e.g., \citealt{pedregosa11}). In all of these cases ML requires large training datasets in order to reach a good performance for new, previously unseen, examples.

Unsupervised ML algorithms are becoming more common as well. An unsupervised algorithm takes as an input only feature values, without labels, with the general goal of learning complex relationships that exist in the dataset. In astronomy, unsupervised ML is used for many different tasks, including outlier detection \citep[e.g,][]{protopapas06, meusinger12, nun16, baron17a}, visualization \citep[e.g,][]{gianniotis16, polsterer16, reis18}, dimensionality reduction \citep[e.g,][]{boroson92, kugler15, gianniotis16, polsterer16}, clustering \citep[e.g,][]{zhang04a, baron15}, object retrieval \citep[e.g,][]{reis18a}, and denoising \citep[e.g,][]{schawinski17}.

While proven to be very useful in astronomy, many ML algorithms were not designed for astronomical datasets that can have different uncertainties for different features or objects.  The performance of ML algorithms depends strongly on the signal to noise ratio (SNR) of the objects in the sample, suggesting that the algorithms are affected by the noise. Therefore, including information regarding the noise is expected to improve the overall performance of such algorithms.  Indeed the inclusion of uncertainties in various ML algorithms has been discussed in recent statistical literature \citep[e.g.][]{schennach16,czarnecki13,sexton08,loustau2015}. However, the current implementation of most off-the-shelf ML algorithms, and specifically the ones that are  used in astronomy, does not support analysis of the uncertainties.

In some specific cases, ML algorithms can handle noise in the dataset. This is because some ML algorithms rank the different features according to their relevance to the task at hand, and give a larger weight during the learning process to the most relevant features. Thus, a noisy feature, which is a feature that is poorly measured for many objects, will be ignored by the algorithm during the training process, since it does not carry relevant information. That is, in simple cases where there exist a correlation between the measurements quality and information content, an ML algorithm will ignore noisy features since they do not carry information. However, ML algorithms are less likely to learn more 'complex' noise, such as different objects with different poorly-measured features. We argue that, while this is far from being a common case, for 'complex' enough noise, the information contained in the uncertainties is vital and could not be compensated for, when using measurement values only, even by large amounts of data and computational resources. 

Though statistical literature on the topic of data with uncertainties exist in astronomy, a rather common way to introduce the information of the uncertainties to an ML algorithm is using the uncertainty values as additional features \citep[e.g,][]{bloom12,dlsanto18b}. Since the algorithm is not provided with the explicit statistical relation between the feature value and its corresponding uncertainty, this method is indirect, and the algorithm will not make the most of the information carried by the uncertainties. Examples of direct use of uncertainties in ML applications include \citet{naul18}, who used a recurrent neural network for classification of variable stars. The neural network architecture chosen in this work allowed an explicit use of measurement uncertainty in the loss function. Another example is \citet{das18} who used a Bayesian neural network for estimating stellar distances and ages. A Bayesian neural network learns a probability density function (PDF) instead of a value for each of its model parameters.  Finally, \citet{castro18} created bootstrapped samples of the features in their dataset, and by using a bagging approach, diminished the effect of feature variance on the classification performance. 

In this work we focus on the commonly used Random Forest algorithm \citep[][]{breiman01}, and modify it to properly treat measurement uncertainties. RF is simple to use and shows high performance for a wide variety of tasks, making it one of the most popular ML algorithms in astronomy. RF is mainly used as a supervised algorithm for classification and regression  \citep[e.g,][]{carliles10, bloom12, pichara12, pichara13, moller16, miller17, plewa18, yong18}, but can also be used for unsupervised learning \citep[e.g, ][]{baron17a, reis18a, reis18} by learning distances between the objects in the sample \citep{breiman03, shi06}. In this work, we focus the discussion on the RF classifier, but the method we present can be easily generalized to other RF use cases.

In Section \ref{sec:RF} we review the original RF algorithm and in Section \ref{sec:PRF} we describe the Probabilistic Random Forest (PRF) algorithm. We compare the performance of the PRF and the original RF in Section \ref{sec:exp}. We discuss our results in Section \ref{sec:dis}, and conclude in Section \ref{sec:sum}. We make our Python implementation of the PRF publicly available at \footnote{\href{https://github.com/ireis/PRF}{https://github.com/ireis/PRF}}.

\section{The original Random Forest algorithm}\label{sec:RF}

PRF is a modified version of the original RF algorithm. We therefore start by describing key aspects of the original RF algorithm in this section. We then present the modifications we made to the original RF algorithm to enable uncertainty treatment in Section \ref{sec:PRF}.

RF is an ensemble learning method, that operates by constructing a large number of decision trees during the training process \citep{breiman01}. A decision tree is a non-parametric model, which is described by a tree-like graph, and is used in both classification and regression tasks. In a decision tree, the relation between the features and the target variable is represented by a series of conjuncted conditions that are arranged in a top-to-bottom tree-like structure. Each condition is of the form: $x_j > x_{j,th}$, where $x_j$ is the value of the feature at index $j$, and $x_{j,th}$ is the threshold, where both the feature and the threshold are determined during the training process. 

To describe how such a tree is constructed we consider the case of a classification task with two classes. The training process starts with the full training dataset and a single tree node, which is called the root of the tree. The algorithm searches for the 'best split', which is a combination of a feature and a threshold that will result in the 'best' separation between the objects of the two classes. The definition of 'best' is a parameter of the algorithm, with a common choice being the \emph{Gini impurity}. The Gini impurity of a group is the probability that a randomly-selected object will be misclassified, if it is assigned with a label that is randomly-drawn from the distribution of the labels in the group. If $P_{n,A}$ and $P_{n,B}$ are the fractions of objects of classes $A$ and $B$ within the group in the node $n$ (also called class probabilities), the Gini impurity $G$ is:
\begin{equation}\label{eq:gini}
\mathrm{G} = 1 - (P_{n,A}^2 + P_{n,B}^2).
\end{equation}
The algorithm iterates over the available features and all possible thresholds. For each threshold, the training data is divided into two groups, \emph{right} and \emph{left} groups, which consist of objects that are to the right and to the left of the threshold respectively. The algorithm searches for the splitting threshold that results in the minimal combined impurity of the two groups:
\begin{equation}\label{eq:split_gini}
\mathrm{G_{right}} \times f_{\mathrm{right}} + \mathrm{G_{left}}  \times f_{\mathrm{left}}, 
\end{equation}
where $\mathrm{G_{right}}, \mathrm{G_{left}}$ are the Gini impurities of the two groups, and $f_{\mathrm{right}}, f_{\mathrm{left}}$ are the fractions of objects in each group, such that $f_{\mathrm{right}} + f_{\mathrm{left}} = 1$. The condition of the root is set to be the feature and the corresponding threshold that result in the minimal combined impurity.

The training data is then split into two groups according to the condition in the root. Objects that satisfy the condition propagate to the \emph{right} node, and objects which do not satisfy the condition propagate to the \emph{left} node. In each of these nodes, the algorithm searches for the 'best split' for the objects that propagated to it. This process is repeated recursively from top to bottom, resulting in a tree-like structure. This process is repeated as long as the two groups have a lower combined impurity compared to the impurity of the node. The nodes which do not satisfy this are called terminal nodes or leafs, and they are the end of their corresponding tree branch. Such nodes are assigned a probability for each of the two classes, according to the distributions of the objects that reached them. Thus, the training process creates a tree-like structure, where the regular nodes contain a condition, and the terminal nodes contain a class value. 

To predict the class of an unlabeled object with a decision tree, the object is propagated through the tree according to its feature values and the conditions in the nodes, until reaching a terminal node. The predicted class of the unlabeled object is the class with the highest probability within the terminal node. In its simplest version, there are no restrictions on the number of nodes and the depth of the resulting decision tree. This results in a classifier with a perfect performance on the training set, and a poor performance on new, previously unseen, datasets. A single decision tree is prone to overfitting the training data, and cannot generalize to new datasets \citep{breiman01}.

RF is a set of many decision trees. The randomness of an RF is induced by: (1) training the different decision trees on randomly-selected subsets of the full dataset, and (2) using random subsets of the features in each node of each decision tree. This randomness reduces the correlation between the different decision trees, resulting in trees with different conditions in their nodes and different overall structures. The RF prediction is an aggregation of the decision trees in the forest, by means of a majority vote. That is, an unlabeled object is propaged through all the trees in forest, where each tree provides a prediction of the object's class, and the final prediction is the label reached in the majority of trees. Furthermore, the fraction of the trees that vote for the predicted class serve as a measure of certainty of the resulting prediction. While a single decision tree tends to overfit the training data, the aggregation of many decision trees in the form of an RF was shown to generalize well to previously unseen datasets, resulting in a better performance \citep{breiman01}.

\section{Probabilistic Random Forest}\label{sec:PRF}

PRF, that we present here, is an RF-based classification algorithm, designed to improve the predictive capabilities of the traditional RF. This is done by accounting for uncertainties in the input data and utilizing their information contents.

Qualitatively, RF receives a sample of observed random pairs, $$(x_1,y_1),...,(x_n,y_n),$$that obey some unknown relation, $h: X\rightarrow{}Y$.  This relation is modeled, to some degree of confidence, then used to predict $y$ for a given value of $x$. PRF, on the other hand, receives a sample of quadruplets, $$(x_1,y_1,\Delta x_1,\Delta y_1),...,(x_n,y_n,\Delta x_n,\Delta y_n),$$
in which $\Delta x_k$ and $\Delta y_k$ represents the uncertainty in the measurements. The PRF will model the relation $h$, while considering the accuracy of each measurement. The modeled relation will predict $y$ for a given pair $(x, \Delta x)$.

The input to a supervised classification task consists of two uncertainty types. (i) Feature uncertainty, $\Delta x$, e.g., uncertainty of a photometric measurement. (ii) Label uncertainty, $\Delta y$ , i.e., uncertainty in the object-type classification in the training stage. Label uncertainties can originate, for example, from human voting-based classification, used by projects such as the Galaxy Zoo \citep{lintott08}. Naturally, the output of a supervised classification task consists only of uncertainties of the latter type.

To account for the uncertainties in the data we treat the features and labels as probability distribution functions, rather than deterministic values. The PRF features become PDFs with expectancy values that are equal to the supplied feature values, and the variances are equal to the corresponding uncertainties squared. The labels become probability mass functions (PMFs); that is, each object is treated as having each label assigned to it with some probability. The differences between the PRF and the regular RF follow naturally from this treatment, as in the limit of negligible uncertainties the PRF converges to the original RF.

\begin{figure}
  \begin{center}
  \includegraphics[width=\columnwidth]{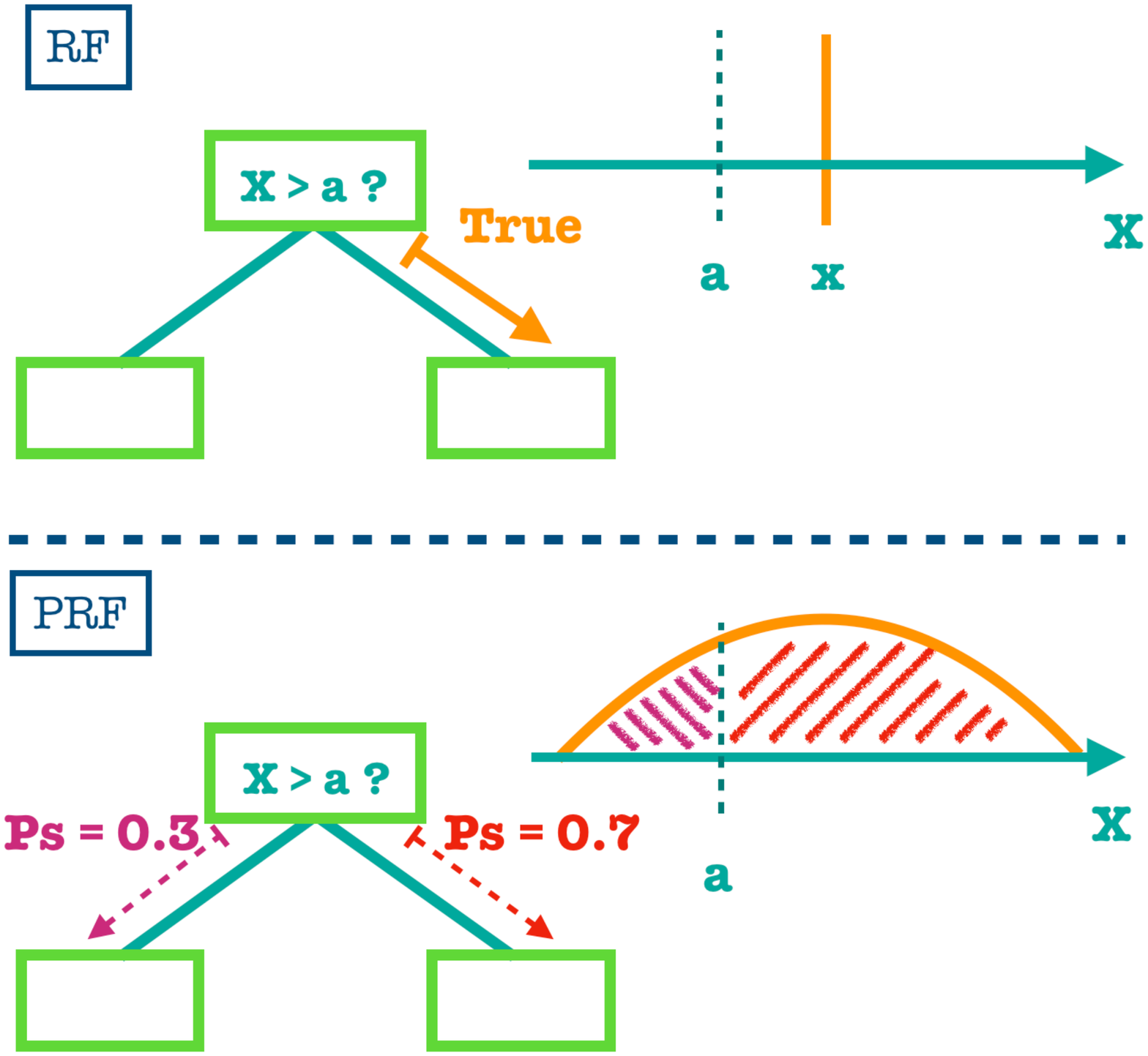}
  \caption{An illustration of the object propagation procedure in the tree nodes, in PRF vs. regular RF. In both algorithms each node of the tree consists of a condition on a specific feature value. In a regular RF, an object is propagated to either the right or left branch according to the outcome of the node's condition for the object's feature value (True $=$ right branch, False $=$ left branch). In the PRF, split probabilities ($\pi_{i}(r)$ and $\pi_{i}(l)$) for the outcome of the node's condition are calculated using the uncertainty in the object feature value. The object is propagated to both branches.}\label{fig:node}
  \end{center}
\end{figure}

\begin{figure*}
  \begin{center}
  \includegraphics[width=\textwidth]{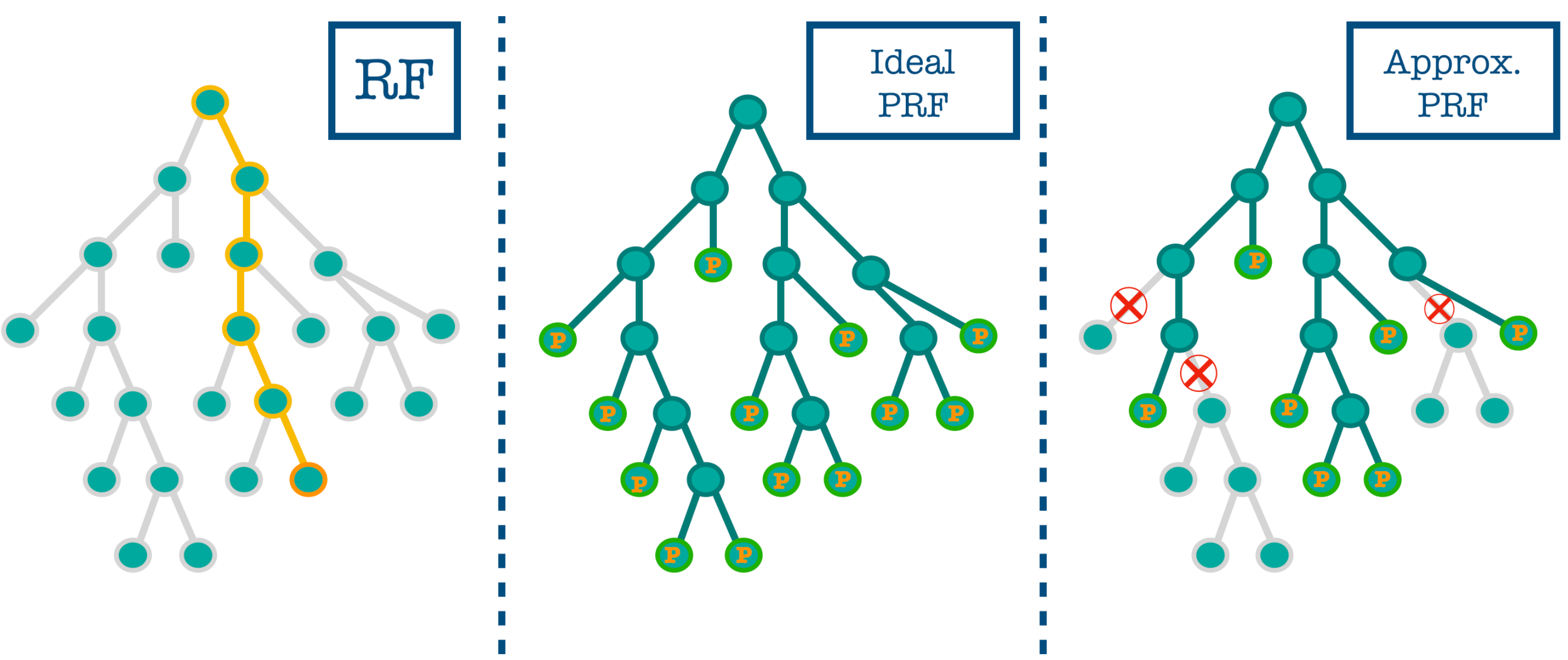}
  \caption{Illustration of the propagation of an object in an RF, ideal PRF, and approximate PRF tree. The left panel shows the propagation of a single object in a regular RF tree. The object follows a single trajectory (the highlighted nodes in the figure) according to the binary criteria in the different nodes. The middle panel shows the propagation of the same object in an ideal PRF tree. The object is propagated to all branches, with the probabilities calculated and stored along the way. In the right panel the propagation of the object in our PRF implementation is shown. This panel shows the influence of the probability threshold parameter on the PRF algorithm (see Section \ref{sec:OPRF}). In this case branches for which the probability drops below the threshold are discarded. These branches are marked with X's in the figure. This results in the algorithm taking only the high probability branches into account, reducing the runtime.}\label{fig:pth}
  \end{center}
\end{figure*}

Let us first consider the effect of feature-uncertainty. In a classical decision tree, each object propagates to either the right or the left branch of each tree node according to the splitting criterion enforced. However, in the presence of feature uncertainty the split is not deterministic. That is, at every node each object may propagate into both branches with some probability. This is illustrated in Figure \ref{fig:node}. Therefore, while in an RF tree an object goes through a single trajectory, in a PRF tree the object may propagate through all the tree nodes with some probability. The propagation down a tree in the two cases is illustrated in the left and middle panels of Figure \ref{fig:pth}.

Now, consider the effect of label-uncertainty. In the presence of label-uncertainty, the classification of an object is described by a PMF. The randomness induced by this PMF differs from the RF user-injected randomness, since it is not drawn from a known, predefined, distribution but rather from some underlying physical or observational source with information content that may be relevant for the classification task. This label-uncertainty propagates through the splitting criterion (e.g., \emph{Gini impurity}), and into the predictive model during the construction of the tree. 

While accounting for uncertainties is clearly tractable from a statistical point of view, it may be computationally intensive. As a first step we will present an 'ideal' PRF model, disregarding the obvious obstacles of complexity and runtime. Then we will extend the discussion to describe our general PRF implementation, designed to provide better accuracy within reasonable, user defined, runtime limitations.

\subsection{A toy model---'ideal' two-class PRF}\label{sec:ideal_PRF}
First we present an idealized PRF algorithm, disregarding runtime limitations. To further simplify the discussion, we will describe a PRF designed to differentiate between two object classes, A and B. In this case, the classification of the $i$-th object in the training set is a Bernoulli random variable, $C_{i}$, that is distributed according to
\begin{equation}\label{eq:3}
p_{i, A} \equiv \mathrm{Pr}(C_i = A) = 1 - \mathrm{Pr}(C_i = B).
\end{equation}
Each object, training set member or otherwise, also has a set of features that were measured with some uncertainty. In the PRF these values represent some underlying distribution. For simplicity, we assume here  that the features are all normally distributed. The algorithm itself, however, is not restricted can  any specific PDF. The distribution of the $j$-th feature in the $i$-th object, $X_{i,j}$, is 
\begin{equation}\label{eq:4}
X_{i,j} \sim \mathcal{N}(x_{i,j},\,\Delta x_{i,j}^{2}),
\end{equation}
where $x_{i,j}$ is the measured feature value and $\Delta x_{i,j}$ is its corresponding uncertainty. In practice, the cumulative distribution function,
\begin{equation}\label{eq:5}
F_{i,j}(\chi) \equiv \mathrm{Pr}(X_{i,j} \leq \chi) = 1 - \mathrm{Pr}(X_{i,j} > \chi),
\end{equation}
is of greater interest in our analysis and will be used in the following.

\paragraph{(i) Propagation in a probabilistic tree}
As a first step, we assume that all the splitting criteria are determined. Specifically, let the first split at the top node be based on the $k$-th feature; if $x_{\dot,k} < \chi_{1}$ the object propagates rightwards. The $i$-th object may now propagate from the top node to the \emph{right} with probability $\pi_{i}(r)$, where:
\begin{equation}\label{eq:6}
\pi_{i}(r) = F_{i,k}(\chi_{1}),
\end{equation}
or to the \emph{left} with probability
\begin{equation}\label{eq:7}
\pi_{i}(l) = 1 - F_{i,k}(\chi_{1}).
\end{equation}
Now, consider a node, $n$, that is located deep in the tree. The probability of the $i$-th object to reach this node is the combined probably for it to take all the turns that led to it from the top node. For example, the probability for the $i$-th object to propagate from the top node twice to the right and then once to the left $(n=r,r,l)$ is
\begin{equation}\label{eq:8}
\pi_{i}(r,r,l) = F_{i, k_{1}}(\chi_{1}) \times F_{i, k_{2}}(\chi_{2}) \times \big(  1 - F_{i, k_{3}}(\chi_{3}) \big).
\end{equation}
This result can be generalized to any arbitrary node, 
\begin{equation}\label{eq:9}
\pi_{i}(n) = \prod_{\eta \in R} F_{i, k_{\eta}}(\chi_{\eta}) \times \prod_{\xi \in L} \big( 1 - F_{i, k_{\xi}}(\chi_{\xi}) \big),
\end{equation}
where $R$ and $L$ are the sets of right and left turns, respectively.

\paragraph    {(ii) Training of a probabilistic tree}
A top-to-bottom construction of a probabilistic tree begins with a set of objects. Each object has an assigned label and a set of measured features. At each tree node, the dataset is split into two subsets, \emph{left} and \emph{right}, such that the two resulting subsets are more homogeneous than the set in their parent node. At each node the algorithm searches for the split that will produce two nodes, more homogeneous than their parent node. To do so, one must define a cost function that will be minimized by the algorithm. In this work we have used a modified version of \emph{Gini impurity}, described in Section \ref{sec:RF}. 

The transition from classical to probabilistic tree affects the ability to determine the node impurity. This is because the class probability, fraction of objects in some given class, is now a random variable by itself. As a consequence, instead of calculating the fraction of objects in the node $n$, we use its expectancy value, 
\begin{equation}\label{eq:10}
P_{n, A} \rightarrow \bar{P}_{n, A}	= \frac{\sum_{i \in n} \pi_{i}(n) \cdot p_{i, A}}{\sum_{i \in n} \pi_{i}(n)},
\end{equation}
\begin{equation}\label{eq:11}
P_{n, B} \rightarrow \bar{P}_{n, B}	= \frac{\sum_{i \in n} \pi_{i}(n) \cdot p_{i, B}}{\sum_{i \in n} \pi_{i}(n)}.
\end{equation}
The modified \emph{Gini impurity} is transformed to 
\begin{equation}\label{eq:12}
G_{n} \rightarrow \bar{G}_{n} = 1 - \big( \bar{P}_{n, A}^{2} + \bar{P}_{n, B}^{2} \big).
\end{equation}
Following the prescription given in Section \ref{sec:RF}, we define the cost function as the weighted average of the modified impurities of the two nodes,  
\begin{equation}\label{eq:13}
\bar{G}_{(n,r)} \times \frac{\sum_{i \in (n, r)} \pi_{i}(n, r)}{\sum_{i \in n} \pi_{i}(n)} + \bar{G}_{(n,l)} \times \frac{\sum_{i \in (n, l)} \pi_{i}(n, l)}{\sum_{i \in n} \pi_{i}(n)}.
\end{equation}
The modified propagation scheme and cost functions are the two major conceptual changes that separate the PRF from the RF. The manner in which the best split of a node is searched for in our PRF implementation is described in the appendix. Other details on the training, such as the split stopping criterion, do not differ from the classical RF. 

\paragraph    {(iii) Prediction by a probabilistic tree and the transition to PRF}
Once a tree is built, it can be used to classify unlabeled objects. The propagation of these objects down the tree is identical both during the tree training and prediction stages, and described by equations (\ref{eq:6}--\ref{eq:13}). Once an object has reached a terminal node, the class probability can be used to provide the prediction of that node, as in the classical RF. However, in a probabilistic tree, each object reaches all the terminal nodes with some probability. One must account for all the predictions given by all the terminal nodes to obtain the prediction of the tree, that is 
\begin{equation}\label{eq:14}
\mathrm{Pr}(A) = \sum_{terminal\,nodes} \pi(n) \times \bar{P}_{n,A},
\end{equation}
\begin{equation}\label{eq:15}
\mathrm{Pr}(B) = \sum_{terminal\,nodes} \pi(n) \times \bar{P}_{n,B}.
\end{equation}
We now discuss the transition from a single tree prediction, to that of the ensemble. In the RF implementation by {\fontfamily{cmtt}\selectfont scikit-learn}\footnote{\href{http://scikit-learn.org/stable/}{scikit-learn.org/stable/}} version 0.19.2 \citep{pedregosa11}, the prediction of the ensemble is carried out via a majority vote. For the PRF predictive model to converge to that of an RF, as the input uncertainties approach zero, the PRF prediction is also determined via majority vote, where the vote of each tree is the label that achieved the largest probability in that tree.

\subsection{A general PRF implementation}\label{sec:OPRF}
\paragraph    {(i) Multi-class PRF}
The general classification task may consist of more than two labels. Each object in the training is assigned with a label probability that follows some general probability mass distribution. That is, the $i$-th object is labeled according to
\begin{equation}\label{eq:16}
\mathrm{Pr}(C_{i} = X) =
    \begin{cases}
   	p_{i, A_{1}} \: X = A_{1}, \\
    \quad\vdots \\ 
    p_{i, A_{k}} \: X = A_{k}.
    \end{cases}
\end{equation}

where $p_{i,A_{1}} + \cdot \cdot \cdot + p_{i, A_{k}} = 1$. Obviously, the number of class probabilities, $\bar{P}_{n, *}$ equals to the number of classes. The modified \emph{Gini impurity} becomes
\begin{equation}\label{eq:17}
\bar{G}_{n} = 1 - \sum_{j=1..k} \bar{P}_{n,A_{j}}.
\end{equation}

\paragraph    {(ii) Selective propagation scheme}
Propagating all objects to all branches of the tree, as done in the ideal PRF algorithm, may require a considerable amount of computation time. However, for a given object there may be nodes to which the propagation probability is small. Pruning the tree at these nodes will reduce the algorithms runtime without impairing its performance. We therefore introduce the probability threshold, $p_{th}$. This tunable parameter sets a lower limit for object propagation, 
\begin{equation}\label{eq:18}
\pi(n) > p_{th},
\end{equation}
that must be fulfilled in order for an object to propagate to $n$. This selective propagation scheme is illustrated in the right panel of Figure \ref{fig:pth}, where a given object does not propagate to branches with a low probability.

The PRF prediction is preformed on the terminal nodes, therefore each object must propagate to at least one terminal node during both the training and prediction stages. Therefore, an exception to the criterion in equation (\ref{eq:18}) is made for objects that do not reach any of the terminal nodes. In this case, the terminal node with the largest $\pi(n)$ is chosen as the single trajectory of the given object. In the limiting case where $p_{th}=0$, each object propagates to all the terminal nodes, as in an ideal PRF. On the other hand, if $p_{th}=1$, each object propagates to a single terminal node as in classical RF.

\begin{figure}
  \begin{center}
  \includegraphics[width=\columnwidth]{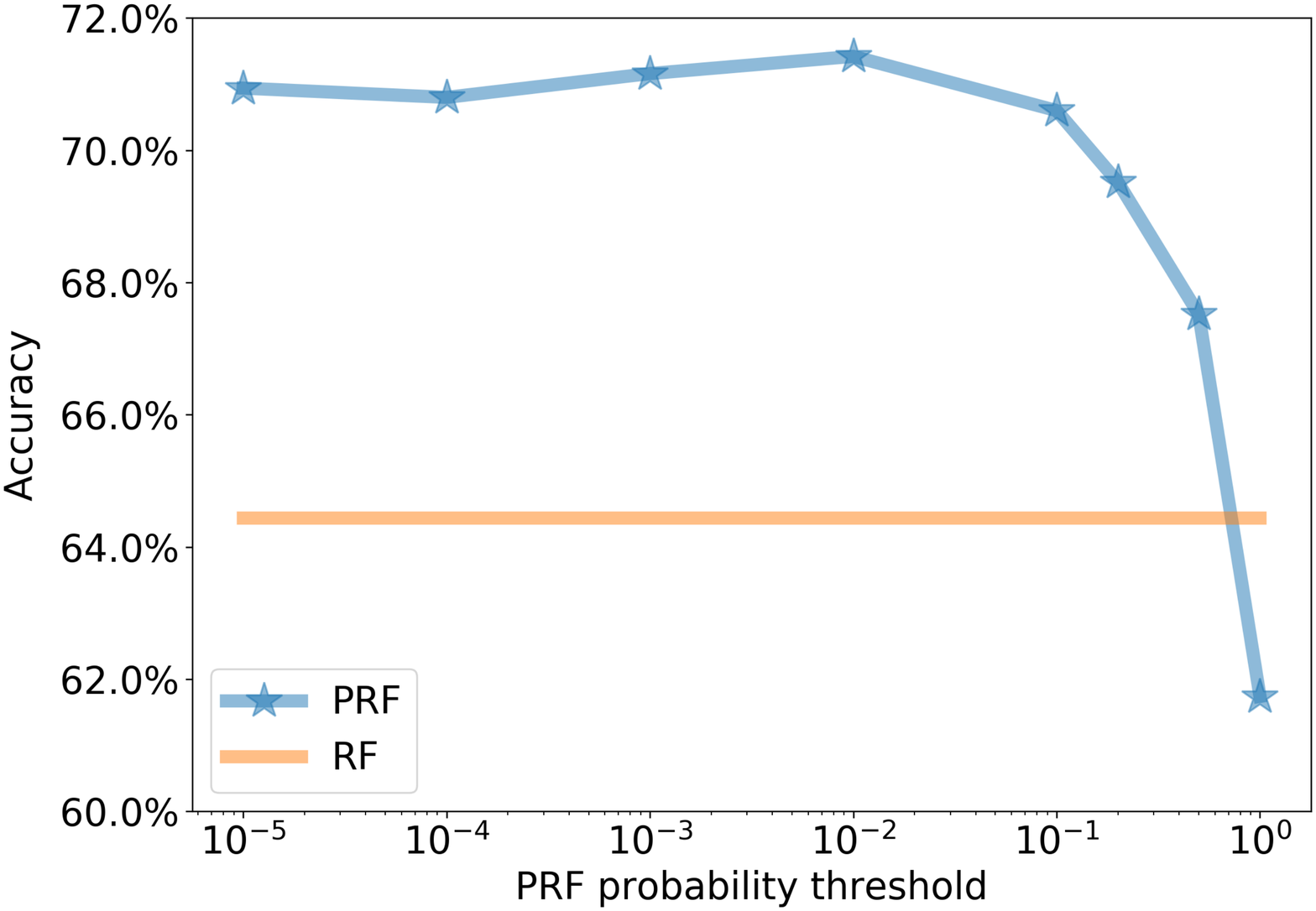}
  \caption{An example for the dependance of the PRF classification accuracy on the probability threshold parameter, $p_{th}$. Our PRF implementation is an approximation for the ideal PRF algorithm due to computational considerations, and $p_{th}$ sets the approximation level. Decreasing the probability threshold makes our PRF implementation slower but more similar to the ideal PRF. The example presented in this figure shows that the accuracy saturates below some value. We found that a value of $p_{th} = 5\%$ is usually sufficient to reach convergence.}\label{fig:keep_proba}
  \end{center}
\end{figure}

Numerical experiments in this work were performed by taking $p_{th}=5\%$. We found that reducing this threshold does not improve the prediction accuracy. Figure \ref{fig:keep_proba} shows an example of the effect that $p_{th}$ has on the prediction accuracy. One can see that below a certain probability threshold, the accuracy converges to a constant value. The impact of $p_{th}$ on the PRF runtime is described in the appendix.

\section{Experiments}
\label{sec:exp}
To test the performance of the PRF, and compare it to RF, we constructed a synthetic dataset and injected various types of noise to it. We created synthetic classification data using {\fontfamily{cmtt}\selectfont scikit-learn}\footnote{\href{http://scikit-learn.org/stable/}{scikit-learn.org/stable/}} version 0.19.2 \citep{pedregosa11}. We constructed a dataset with two classes and 15 features, out of which 10 features are informative. We used 5,000 objects for the training set and 5,000 objects for the test set. The choice of the dataset, namely the number of features and the size of the training and test samples, is arbitrary, and the same qualitative behaviour is observed for different datasets. In order to examine the robustness of the PRF to noisy datasets, we consider four different types of noise: 
\begin{enumerate}
\item{\textbf{Noise in the labels}, where a fraction of the objects in the training set are misclassified.}
\item{\textbf{Noise in the features (simple case)}, where the dataset consists of features that are well-measured and features that are poorly measured.}
\item{\textbf{Noise in the features (complex case)}, where different subsets of the dataset consist of different poorly-measured features.}
\item{\textbf{Different noise characteristics in the training and the test sets.}}
\end{enumerate}
Throughout the experiments, we compare the classification accuracy of the PRF to the {\fontfamily{cmtt}\selectfont scikit-learn} version 0.19.2 implementation of the RF on the test set. The noisy datasets and all the experiments described below are publicly-available at \footnote{\href{https://github.com/ireis/PRF}{https://github.com/ireis/PRF}}.

\paragraph    {(i) Noise in the labels}
First, we consider the case in which there are uncertainties in the labels but not in the feature values. We inject noise to the labels in the following way. For each object, a probability for switching its label, $p_{\mathrm{switch}}$, is randomly drawn from a uniform distribution between 0 and 0.5. Then, the label of each object is randomly switched with a probability of $p_{\mathrm{switch}}$. That is, the label remains the same with a probability of $1- p_{\mathrm{switch}}$. The synthetic dataset with the noisy labels serve as the input to the RF algorithm. The PRF takes as an input both the noisy labels and their corresponding $1- p_{\mathrm{switch}}$ probabilities, where the latter represent the uncertainty in the input labels. This experiment can be thought of as having a device that measures the labels, where it is known that the device lies $p_{\mathrm{switch}}$ percent of the time. This type of noise is known in the ML literature as random classification noise \citep[][]{angluin88}.

\begin{figure}
  \begin{center}
  \includegraphics[width=\columnwidth]{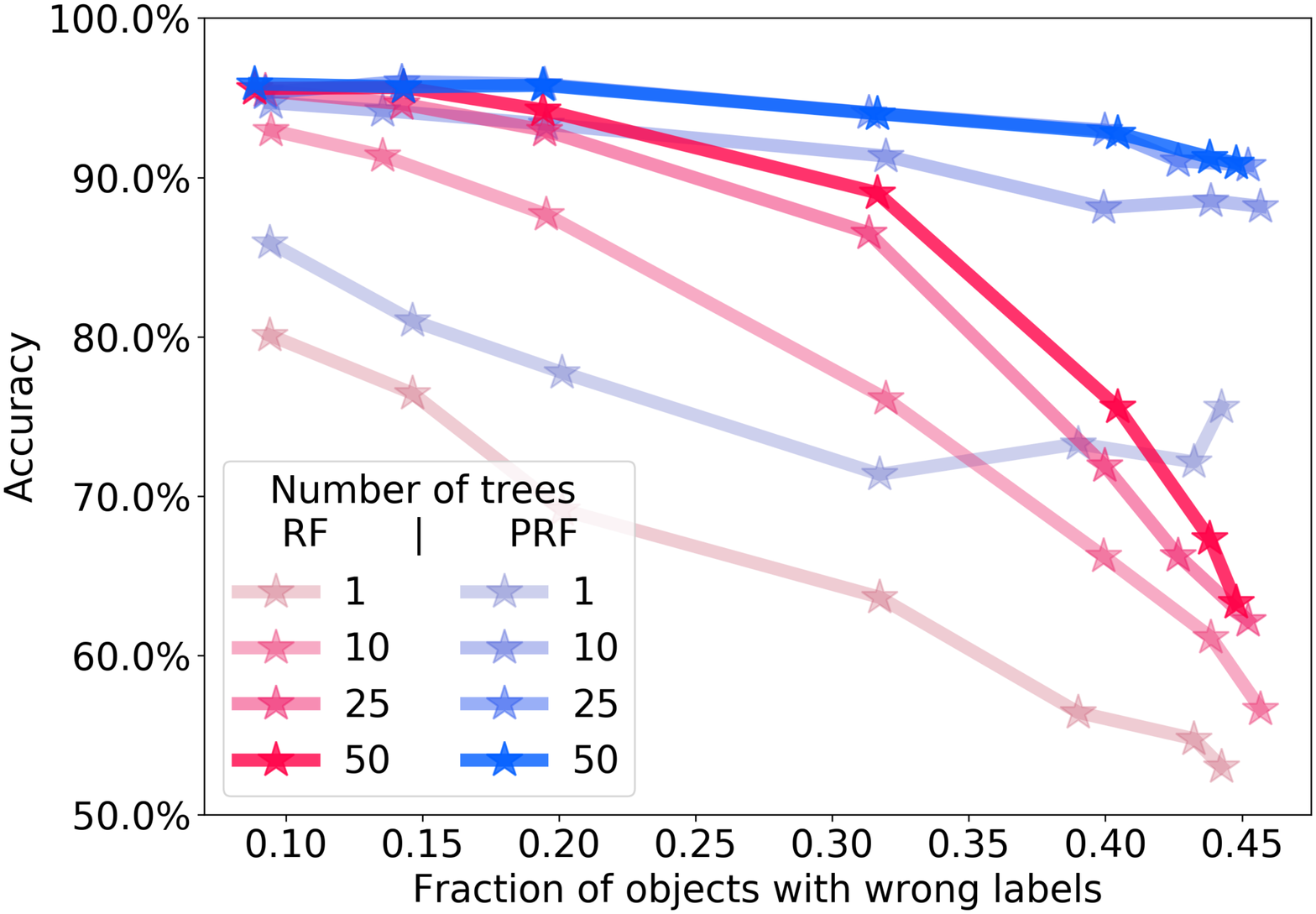}
  \caption{The classification accuracy of the PRF (blue) and the RF (red) versus the fraction of wrong labels in the dataset, for different number of trees. The fraction of wrong labels is defined as the number of objects for which the label was randomly switched. We consider 1, 10, 25, and 50 trees in the forests.}
  \label{fig:noise_in_labels}
  \end{center}
\end{figure}

In Figure \ref{fig:noise_in_labels} we show the classification accuracy of the algorithms versus the fraction of bad labels in the dataset, for different number of trees. The number of bad labels is defined as the number of objects for which their label was randomly switched. We show the accuracies of the algorithms for 1, 10, 25, and 50 trees, where we mark the classification accuracy of the RF with red lines, and the accuracy of the PRF with blue lines. As expected, the classification accuracy of both of the algorithms increases with the number of trees in the forest, until reaching convergence. Second, the classification accuracy generally decreases with the fraction of bad labels in the dataset. We point out the robustness of the original RF to noisy labels: one can see that its classification accuracy drops by less than 10\% for a dataset with as many as 30\% wrong labels. Finally, Figure \ref{fig:noise_in_labels} shows that the PRF outperforms RF in all cases. The PRF accuracy converges faster as a function of the number of trees in the forest, and shows a decrease of less than 5\% for a dataset with more than 45\% wrong labels. Since the original RF accuracy converges to a lower value than that of the PRF, increasing the number of trees in the RF will not lead to a performance that is similar to the PRF.
 
We note that unlike the experiments discussed below, in this experiment we inject noise only to the training set. This is done in order to estimate the true accuracy of the algorithms, since the accuracy on a test set with noisy labels will just reflect the fraction of bad labels in the sample. This is not possible, of course, in real-life applications, where the test set accuracy is determined by both the quality of the labels and the algorithm performance. In such cases, in order to estimate the true accuracy of the algorithm, we suggest choosing a sub-sample of the dataset, for which the label uncertainties are negligible, and use this as the test set. As shown in Figure \ref{fig:noise_in_labels}, contrary to the test set, the training sample can contain objects with uncertain labels, since the PRF is robust to noisy labels.

\paragraph    {(ii) Noise in the features - random noisy objects and random noisy features}
In this experiment we add noise to the feature values, and not to the labels. The noise has a Gaussian distribution, and its magnitude is set randomly for each object and feature in the dataset. We apply the following procedure. A per-object noise coefficient, $N_o$, is randomly drawn from a uniform distribution between 0 and 1 for each object in the dataset. A per-feature noise coefficient, $N_f$, is randomly drawn from a uniform distribution between 0 and 1 for each feature in the dataset. Then, each measurement, which is a specific feature value of a specific object, is assigned with the noise coefficient $N_{o,f} = N_{o} \times N_{f} \times N_{s}$, where $N_{s}$ is the noise coefficient of the entire dataset (training and test set). Throughout the experiment, we vary $N_{s}$ and examine the RF and PRF performance. The uncertainty of a single measurement is defined as $\sigma_{o,f} = N_{o,f} \times \sigma_{f}$, where $\sigma_{f}$ is the standard deviation of the given feature, $f$, over all objects. We multiply the noise coefficient by $\sigma_{f}$ in order to preserve the same physical units in the noisy data. Finally, the noisy measurement is drawn from the distribution $\tilde{x}_{o,f} \sim \mathcal{N} \left(x_{o,f}, \sigma_{o,f}^{2}\right)$, where $x_{o,f}$ is the original measurement, and $\tilde{x}_{o,f}$ is the data after the noise injection. We note that such noise injection naturally results in objects which are measured with a higher precision and objects that are measured with a lower precision. Furthermore, such a procedure naturally results in features which are poorly-measured in the dataset, and features that are well-measured. Therefore, we expect the RF and PRF algorithms to ignore the poorly-measured features, and assign a larger weight to the well-measured features.

The noisy dataset, namely $\tilde{x}_{o,f}$, serve as the input to the RF algorithm. The PRF takes as an input both $\tilde{x}_{o,f}$, and their corresponding uncertainties $\sigma_{o,f}$. To evaluate the RF and PRF performance as a function of the noise level in the dataset, we define $<N_{o,f}>$. $<N_{o,f}>$ is $\sigma_{o,f}/\sigma_{f}$, averaged over the different features and objects. It represents the average scatter due to the noise with respect to the intrinsic scatter of the original dataset, and can be considered as the average inverse of the SNR of the dataset. We note that different noise injection methods will result in very different $<N_{o,f}>$ values, and thus their absolute values are meaningless. Therefore, we focus our attention on the accuracy of the algorithms with increasing $<N_{o,f}>$.

\begin{figure}
  \begin{center}
  \includegraphics[width=\columnwidth]{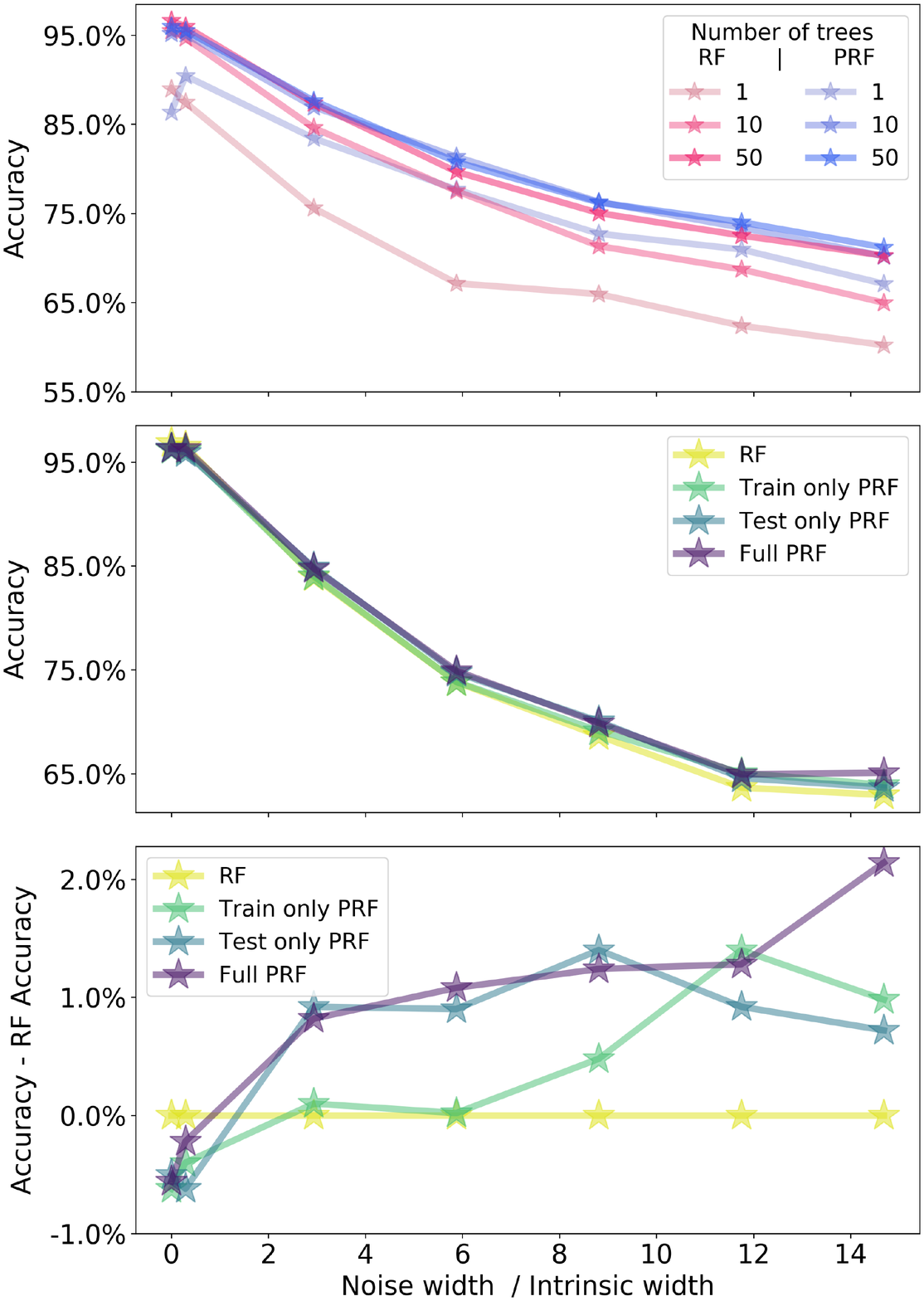}
  \caption{Comparison between the RF and the PRF accuracy for the case of random noisy objects and random noisy features, versus the noise level in the dataset. The injected noise and the definition of the noise level are described in the text. The top panel compares the results obtained with different number of trees. The middle panel shows the RF accuracy and the accuracy of the PRF for three cases: 'training-only' PRF case, where the PRF takes as an input the uncertainties of the training set, but is not provided with the uncertainties in the test set (i.e., these are set to zero), a 'test-only' PRF case, where the PRF takes as an input the uncertainties in the test set, but is not provided with the uncertainties of the training set, and a 'full-PRF' case. The bottom panel shows the difference between these accuracies and the RF accuracy as a function of the noise level in the dataset. The bottom two panels are based on forests with 50 trees.} 
  \label{fig:noise_in_features_white_noise}
  \end{center}
\end{figure}

In the top panel of Figure \ref{fig:noise_in_features_white_noise} we show the RF and PRF accuracy versus the noise level in the dataset. We show the accuracy as a function of the number of trees in the forest, where we use 1, 10, and 50 trees. We mark the RF accuracy with red lines, and the PRF accuracy with blue lines. Similarly to the previous case, the accuracy increases with the number of trees in the forest, until reaching a convergence. In the middle panel of Figure \ref{fig:noise_in_features_white_noise} we compare the accuracy of the RF to the accuracy of the PRF, where we examine several PRF cases. We show a 'training-only' PRF case, where the PRF takes as an input the uncertainties of the training set, but is not provided with the uncertainties in the test set (i.e., these are set to zero). The 'test-only' PRF takes as an input the uncertainties in the test set, but is not provided with the uncertainties of the training set. The 'full-PRF' takes as an input both the uncertainties in the training set and in the test set. We perform this comparison in order to examine which of the modifications that we made to the RF are responsible for the improved performance. Throughout the experiments we perform, we find that both of the modifications are required to reach the best performance. In the bottom panel of Figure \ref{fig:noise_in_features_white_noise} we show the accuracy difference as a function of the noise level in the dataset.

Figure \ref{fig:noise_in_features_white_noise} shows that the classification accuracies of the RF and the PRF decrease as a function of the noise level in the dataset. One can see that the PRF slightly outperforms the original RF, resulting in an accuracy improvement of roughly 1\%. We attribute this minor improvement to the simplicity of the adopted noise. As mentioned above, the noise injection naturally results in features that are generally well-measured in the dataset, and features that are poorly-measured. In this simple case, there exist a correlation between the overall information content and the noise level of a given feature. Therefore, the original RF ignores the noisy features due to their low information content, and a proper treatment of the uncertainties improves the accuracy by a negligible margin. We note that the original RF shows a slightly improved performance, compared to the PRF, for the original dataset (i.e., noise level of zero). Since the PRF is reduced to a simple RF in the case of negligible uncertainties, we attribute this to differences in implementation between our PRF and the {\fontfamily{cmtt}\selectfont scikit-learn} RF version.

\paragraph    {(iii) Noise in the features - groups of objects with different noisy features}
We have seen in experiment (ii) a minor improvement in the classification accuracy of the PRF with respect to the original RF, due to the simplicity of the injected noise. In this experiment, we examine a more complex and realistic noise scenario. We add a Gaussian noise to the feature values in the dataset, using the same procedure outlined in experiment (ii). However, in this case, we divide the dataset into two groups (not according to their labels), and a per-feature noise coefficient $N_f$ is randomly drawn for each group separately. This results in two groups of objects with \emph{different} poorly-measured features, within a single dataset. In this case, there is no simple correlation between the information content and the measurement quality of features throughout the dataset. We therefore expect that a proper statistical treatment of the uncertainties will result in an overall better performance. 

\begin{figure}
  \begin{center}
  \includegraphics[width=\columnwidth]{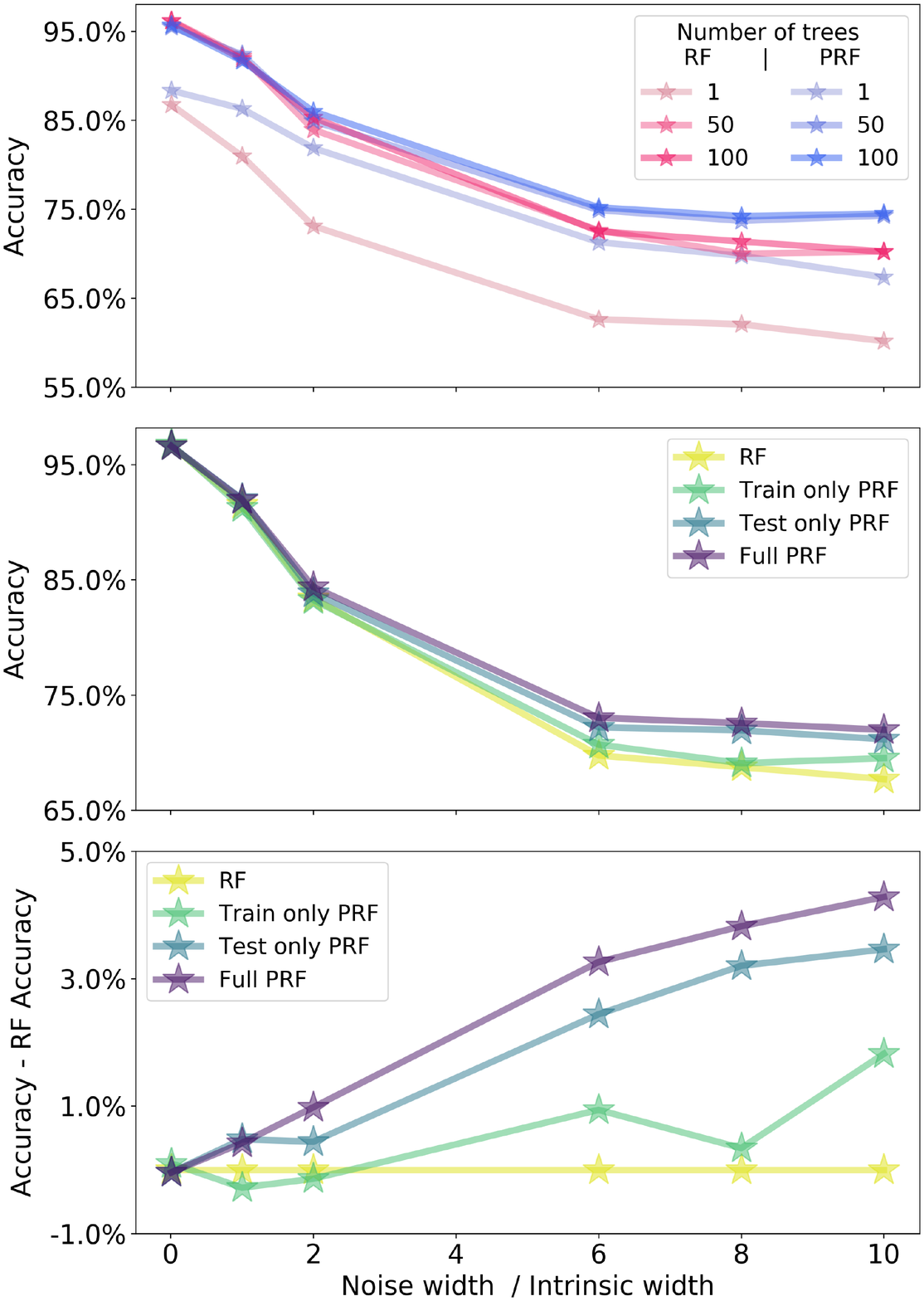}
  \caption{Same as Figure \ref{fig:noise_in_features_white_noise}, but with different noisy features for different groups of objects. The injected noise is described in the text.} 
  \label{fig:noise_in_features_groups}
  \end{center}
\end{figure}

As in the previous case, the original RF takes as an input the noisy dataset, and the PRF takes as an input both the features and their corresponding uncertainties. In Figure \ref{fig:noise_in_features_groups} we compare the performance of the RF to that of the PRF, with similar panels to those we described in experiment (ii). The behaviour of the accuracies is similar to what we found in experiment (ii), but with a larger improvement of the PRF with respect to the original RF. Furthermore, as we divide the original dataset into a larger number of groups, where each group has different per-feature noise coefficients, we find that the PRF outperforms the RF by a larger margin. We therefore find that as the noise becomes more complex, as is usually the case for real measurements, where the noise can depend on a variety of hidden parameters, a proper treatment of the uncertainties allows one to reach better performance. 

\begin{figure}
  \begin{center}
  \includegraphics[width=\columnwidth]{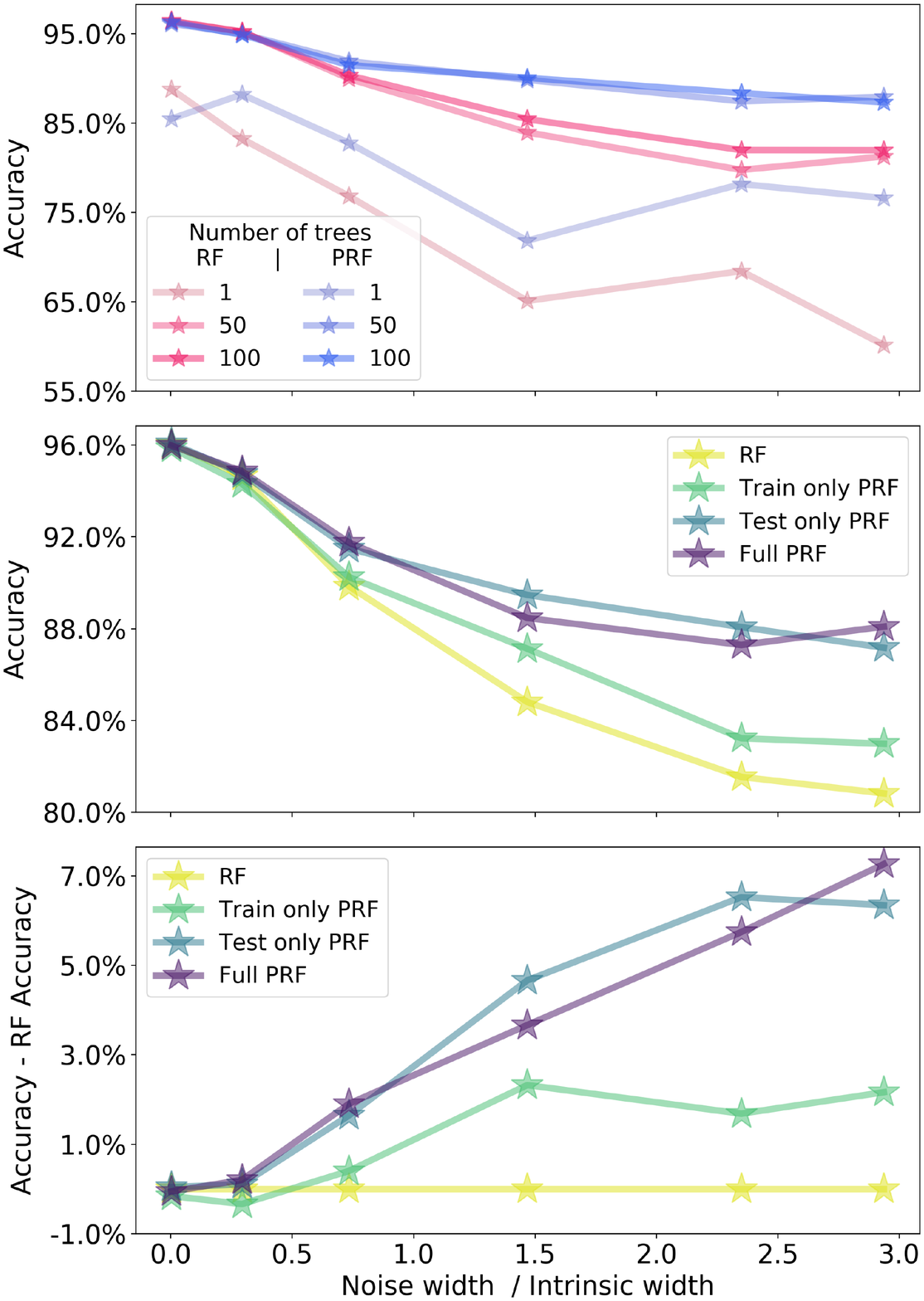}
  \caption{Same as Figure \ref{fig:noise_in_features_white_noise}, but with different noisy features for the training and the test sets. The injected noise is described in the text.} 
  \label{fig:noise_in_features_train_test}
  \end{center}
\end{figure}

\paragraph    {(iv) Noise in the features - different noise for training and test sets}
In this experiment we add Gaussian noise to the feature values in the dataset, using the same procedure outlined in experiment (ii). The difference in this case is that a different per-feature noise coefficient is drawn for the training set and the test set. This results in a training set and a test set with different poorly-measured (and well-measured) features. In practice, such a situation could arise in the case where the training and test sets come from different sources, for example, from two different instruments. As in the previous case, such a dataset does not show a simple correlation between the information content and the measurement quality, and the information obtained during the training process is not directly applicable to the test set, and thus to previously unseen datasets. In Figure \ref{fig:noise_in_features_train_test} we compare the performance of the RF to that of the PRF, with similar panels to those described in experiment (ii). One can see a significant improvement in the classification accuracy when using the PRF, compared to the original RF. Similarly to the previous cases, the PRF accuracy could not be reached by using an RF with a larger number of trees. This has important implications for applications where one wants to learn from one survey and apply to a newer one, a task commonly referred to as 'transfer learning'. If one has a good handle on the uncertainties, PRF can use this information and boost the accuracy. 

\section{Discussion}\label{sec:dis}
The PRF is a modified version of the original RF algorithm, which applies a proper statistical treatment to the data uncertainties. Compared to its precursor, which takes as an input the feature values and the objects' labels, the PRF also takes as an input the feature and label uncertainties. The PRF converges to the original RF algorithm when the uncertainties are set to zero. Throughout the discussion, we assumed that all the noisy measurements have a Gaussian distribution. However, the discussion and our publicly-available code can be easily generalized to other distributions. Moreover, different measurements can be described by different distributions.

We performed several experiments, where we constructed a synthetic dataset with various types of noise. We found that the PRF outperforms the RF in all the cases\footnote{The single exception occurs in experiment (ii) in Section \ref{sec:exp}, where the RF slightly outperforms the PRF for a clean dataset. We attribute this to differences in the implementations of RF and PRF.}, and the difference in classification accuracy between the two increases with increasing noise level and complexity. While the experiments we performed represent simple cases, astronomical datasets are usually complicated, and suffer from heterogeneous noise sources. Since astronomical datasets can be represented by a combination of the scenarios we described in Section \ref{sec:exp}, we argue that the PRF may outperform the RF by an even larger margin when applied to astronomical datasets. Below, we discuss the PRF in the context of other works and methods (Section \ref{sec:related}), and discuss additional properties of the PRF and their implications in Section \ref{sec:prf_props_and_imps}.

\subsection{Related works and methods}\label{sec:related}
\subsubsection{Bootstrapping features}
Recently, \citet{castro18} proposed a novel method to treat feature uncertainties with an ensemble of decision trees. Their dataset consisted of light-curves of variable stars, from which studies typically generate a set of measured features. The typical features used to describe time-series datasets are strongly affected by the quality of the individual measurements, by the cadence of the survey, and by the gaps in the light-curves. The uncertainties of these features are non-trivial, and usually do not have a closed mathematical form. \citet{castro18} used Gaussian processes to obtain a probabilistic description of the observed light-curves. Then, based on this probabilistic model, they bootstrapped samples of time series, from which they generated features. These bootstrapped features served as the input to an ensemble of decision trees algorithm. Therefore, they propagated the uncertainties from the observed light-curves to the final features, representing each measurement as a sample from a complex probabilistic distribution. Their trained model, where uncertainties were taken into account, outperformed the simple RF model which was based only on the measured feature values.

The method presented by \citet{castro18}, namely bootstrapping feature measurements from a PDF, and then training an ensemble of decision trees, is mathematically equivalent to the ideal PRF, when the number of bootstrapped samples approaches infinity. In practice, \citet{castro18} found that 100 bootstrapped samples are enough to reach a convergence in the classification accuracy of their ensemble. There are several similarities and differences between the ideal PRF and the method presented by \citet{castro18}. First, both methods can work with different statistical distributions that describe the feature measurements. However, bootstrapping may become impractical for datasets with a large number of features. In order to sample the PDFs of the features in such a dataset, in may be required to increase dramatically the number of bootstrapped samples, which can become impossible in terms of runtime. Furthermore, representing missing values with the bootstrapping method is a non-trivial task, and may require a large number of bootstrapped samples. Finally, the ideal PRF takes into account both feature and label uncertainties, while \citet{castro18} chose to work only with feature uncertainties, though we note that their method is also applicable to label uncertainties.

\subsubsection{Other ensemble methods}
Ensemble methods rely on creating a diverse collection of classifiers and aggregating their predictions \citep[][]{kuncheva03}. The RF is an ensemble of decision trees, which is generated using a method called 'bagging'. In the 'bagging' method, the training set is split to randomly-selected subsets, and each decision tree is trained on a subset of the data. Since the decision trees are trained on different subsets of the data, the ensemble consists of a diverse population of trees. Another popular ensemble method is 'boosting', for example, the decision tree booting algorithm Adaboost \citep[][]{freund97}. In this method the decision trees are built sequentially, where during training, each decision tree puts more weight on objects that the previous trees failed to classify correctly. 

The modifications that we made to the original RF are at the level of single decision trees. As such, the same modifications can be applied to other ensemble methods, such as decision tree boosting. Boosting algorithms were shown to be more sensitive to noisy objects than bagging algorithms \citep{maclin97, dietterich00}. This is because in boosting, noisy objects have a larger probability to be misclassified by a given decision tree, and thus will be given a larger weight during the training process of the next decision trees in the forest. Thus, in boosting methods, the training process can be dominated by noisy objects, instead of by smaller groups of objects with different intrinsic properties. We argue that the uncertainty treatment we describe in this work is of a larger importance for bootsting algorithms than for bagging algorithms, and we expect that a proper treatment of the measurement uncertainties will significantly improve the performance of boosting algorithms.

\subsection{PRF properties and additional implications}\label{sec:prf_props_and_imps}

\subsubsection{Missing values}
Missing values are common in astronomical datasets. Their presence can be sporadic, where some measurements are not available for a few specific objects, due to different observational or instrumental constrains. Their presence can also be systematic, where a non-negligible subset of the objects in the dataset are missing a certain feature. Most ML algorithms are not designed to handle missing values in the dataset. Usually, when the dataset contains missing values, one of the following procedures is applied. One can disregard objects which have missing values in some of their features, or disregard features which were not measured in all the objects in the sample. Alternatively, the missing values can be replaced by some 'representative' values, such as the mean or the median of the given feature distribution, or via interpolation (see e.g., \citealt{miller17}). The latter procedure always involves some assumptions about the dataset, and can only be done in cases where the missing values constitute a small fraction of the entire dataset. 

Since the PRF treats the feature measurements as PDFs, it can naturally represent missing values, without making assumptions about the dataset. During both the training and the prediction stages, an object with a missing value at a given feature will have a probability of 0.5 to propagate to the \emph{right} node, and a probability of 0.5 to propagate to the \emph{left} node. Thus, we are not required to dismiss objects with missing values prior to the training stage, and the model will be constructed only from the measured features of the given object. For a large enough training set, discarding a few objects will not affect the constructed model significantly. However, this PRF ability allows us to predict the class of unlabeled objects with missing features. Note that while this feature is included in the PRF code we are still investigating additional ways of handling missing values, and the implementation in future versions might change.

\subsubsection{Noise complexity}
\begin{figure}
  \begin{center}
  \includegraphics[width=\columnwidth]{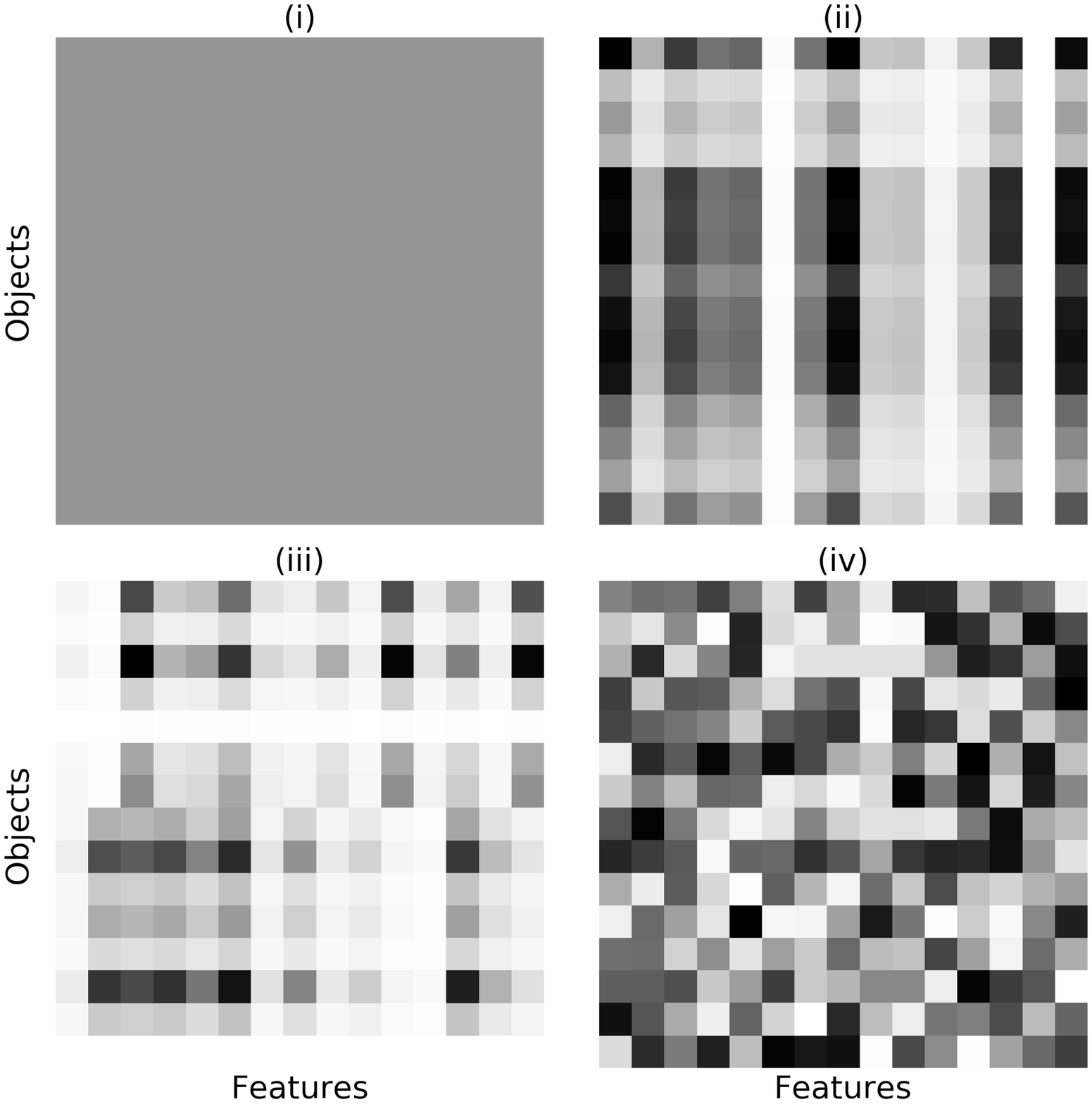}
  \caption{Illustration of the different noise cases discussed in the text. The different panels correspond to different noise cases, and in each panel a point represents the noise strength for a single measurement (that is, a given object and feature). Black represents  a noisy measurements and white a precise one. In case (i) all measurements have the same noise strength. In case (ii) different features are more noisy than others for all objects, and vise versa. In case (iii) two groups of objects have different noisy features. In case (iv) each individual measurement has a different noise strength.} 
  \label{fig:noise_mat}
  \end{center}
\end{figure}

We have shown that taking the uncertainties into account generally results in an improved predictive performance. The level of improvement is largely dependent on the properties of the noise in the dataset. The more complicated the noise is, the more information is contained in the uncertainties, and thus we expect a more significant improvement in the accuracy of the PRF with respect to the original RF. We demonstrate this point with four different noise cases with increasing complexity; (i) The noise level is similar for all the objects and their feature values. In such a case, the noise is described by a single number. (ii) The dataset consists of features which are poorly-measured and features which are well-measured, and of objects with a larger overall SNR and objects with a lower SNR. In this case, the noise can be described by a per-feature and a per-object noise coefficient, that is, by $N_{\mathrm{features}} + N_{\mathrm{objects}}$ numbers. (iii) The dataset consists of several distinct groups, where the noise level in each group is described by different per-feature noise coefficients. Therefore, in this case, the noise can be described by $N_{\mathrm{groups}} \times N_{\mathrm{features}} + N_{\mathrm{objects}}$ numbers. Finally, (iv) every object in the dataset has different poorly-measured features, which is generally described by $N_{\mathrm{objects}} \times N_{\mathrm{features}}$ numbers. The different cases are illustrated in Figure \ref{fig:noise_mat}.

In the very simple cases (i) and (ii), the PRF does not significantly outperform the original RF in terms of classification accuracy. In case (i), since the uncertainties are constant throughout the dataset, including them does not affect the constructed model. In case (ii), since there exist a correlation between the information content and the noise level of each feature, the original RF can ignore the noisy features, since they do not carry information, and a proper statistical treatment of the uncertainties will not improve the model by a large margin. However, as we increase the complexity of the noise, e.g. cases (iii) and (iv), we find that the PRF outperforms the original RF by an increasing margin. This may suggest that, for a given dataset, as the information content of the uncertainties increases (or, the entropy of the noise), it becomes more important to take them into account when constructing a predictive model. A quantitative exploration of these suggestions will be the topic of a future work. 

\subsubsection{Implications to new surveys: label purity and transfer learning}

ML algorithms are usually impractical to use in new surveys. Supervised ML algorithms usually require large training samples with a high level of label purity in order to construct robust models of the dataset. As a result, supervised ML algorithms are used only once a training sample of a sufficient quality was constructed, which takes place long after the survey had started. We examined the performance of the PRF in the presence of noisy labels in experiment (i) in Section \ref{sec:exp}. Strikingly, the classification accuracy of the PRF decreased by less than 5\% for a training set with 45\% wrong labels, compared to a training set with pure labels. This suggests that the PRF is extremely robust against misclassified objects, as long as it is provided with label uncertainties. This opens up a new opportunity to apply supervised learning in new surveys where the quality of objects' labels is poor.

One can, in principle, use a supervised model that was trained on a different survey and apply it to unseen objects in a new survey. This is typically called Transfer Learning. However, many ML algorithms tend to learn not only the intrinsic properties of the objects in the sample, but also the noise characteristics of the dataset. Since different surveys usually have different noise characteristics (e.g., different instruments, calibrations, object selection criteria, etc.), ML algorithms that were trained on a given survey do not generalize well to a different survey. We have shown in experiment (iv) in Section \ref{sec:exp} that the PRF outperforms the RF when applied to a dataset with different noise characteristics in the training and the test sets. This implies that the PRF is a more robust tool to use in Transfer Learning, since it correctly accounts for the different noise characteristics during training and during prediction. 

\section{Summary}
\label{sec:sum}

ML algorithms are gaining increasing popularity in astronomy. They are able to describe complex objects and relations within the dataset, and proved successful in various supervised and unsupervised tasks in astronomy. Despite their notable advantages compared to classical data-analysis methods, most ML algorithms are not designed to take data uncertainties into account. Therefore, the performance of most ML algorithms is highly sensitive to the noise properties of the dataset, and successful ML applications in astronomy are usually based on subsets with high quality measurements. In order to generalize ML algorithms to all astronomical datasets, we must modify existing off-the-shelf tools to apply a proper statistical treatment to noisy measurements.

The Random Forest (RF) algorithm is among the most popular ML algorithms in astronomy and is used in both supervised and unsupervised learning tasks. Despite its excellent performance in astronomical setups, the original RF algorithm does not take into account the data uncertainties. In this work we present the Probabilistic Random Forest (PRF), which is a modified version of the original RF algorithm. The PRF takes as an input the uncertainties in the features (i.e., measurements) and in the labels (i.e., classes), and treats these as probability distribution functions rather than deterministic quantities. This treatment allows the PRF to naturally represent missing values in the dataset, which most algorithms fail to do. Our PRF python implementation is open source, easy to use, and has a similar interface to the RF implementation of {\fontfamily{cmtt}\selectfont scikit-learn}.

To test the robustness of the PRF to noisy datasets, we constructed a synthetic dataset and injected various types of noise to it. We compared the classification accuracy of the PRF to that of the original RF, and found that the PRF outperforms the original RF in most cases. While the PRF requires a longer running time, we found an improvement in classification accuracy of up to 10\% in the case of noisy features. In the case of noisy labels, the PRF outperformed RF by 30\%, and showed a decrease in classification accuracy of less then 5\% when applied to a dataset with as many as 45\% misclassified objects, compared to a dataset with pure labels. This implies that the PRF in extremely robust against misclassified objects, as long as it is provided with label uncertainties, and can be used in new surveys where label quality is still poor. We also found that PRF is more robust than RF when applied to a dataset with different noise characteristics in the training and the test sets, thus it can be used in Transfer Learning tasks.

Our tests revealed that as the noise complexity (i.e., entropy) increases, the PRF outperforms the original RF by an increasing margin. Since real data typically has complicated noise distributions, we find that PRF is more robust than its predecessor and we recommend its use whenever uncertainties can be estimated. Our work can be generalized to other ensemble methods, such as boosting, where we expect an even larger improvement in classification accuracy and overall performance.

\section*{Acknowledgements}
We thank N. Lubelchick, K. Polsterer, P. Protopapas, and D. Poznanski for helpful comments and discussions. We thank the anonymous referee for the insightful comments that helped improving this paper.
This research made use of: {\fontfamily{cmtt}\selectfont  scikit-learn} \citep[][]{pedregosa11}, {\fontfamily{cmtt}\selectfont  SciPy} \citep[including {\fontfamily{cmtt}\selectfont  pandas} and {\fontfamily{cmtt}\selectfont NumPy}][]{scipy01},  {\fontfamily{cmtt}\selectfont IPython} \citep[][]{perez07}, {\fontfamily{cmtt}\selectfont matplotlib} \citep[][]{hunter07},  {\fontfamily{cmtt}\selectfont Joblib} \footnote{\href{https://pythonhosted.org/joblib}{pythonhosted.org/joblib}}, and {\fontfamily{cmtt}\selectfont Numba} \citep{lam15}.

\appendix

\section{Implementation details}
Here we discuss some of the details of our PRF implementation. These include notes about the code itself, the runtime, and implementation choices that do not affect the essence of the algorithm.

\paragraph    {Code notes}
The PRF is implemented entirely in {\fontfamily{cmtt}\selectfont Python}. Wherever possible we used  {\fontfamily{cmtt}\selectfont Numba} \citep{lam15} for just-in-time compilation to native machine instructions, which significantly improved the runtime. As most of the operations are per decision tree, the PRF is easily parallelizable. Both the training and the prediction stages are parallelized at a level of a single decision tree using the {\fontfamily{cmtt}\selectfont Joblib} package. 

\begin{figure}
  \begin{center}
  \includegraphics[width=0.5\columnwidth]{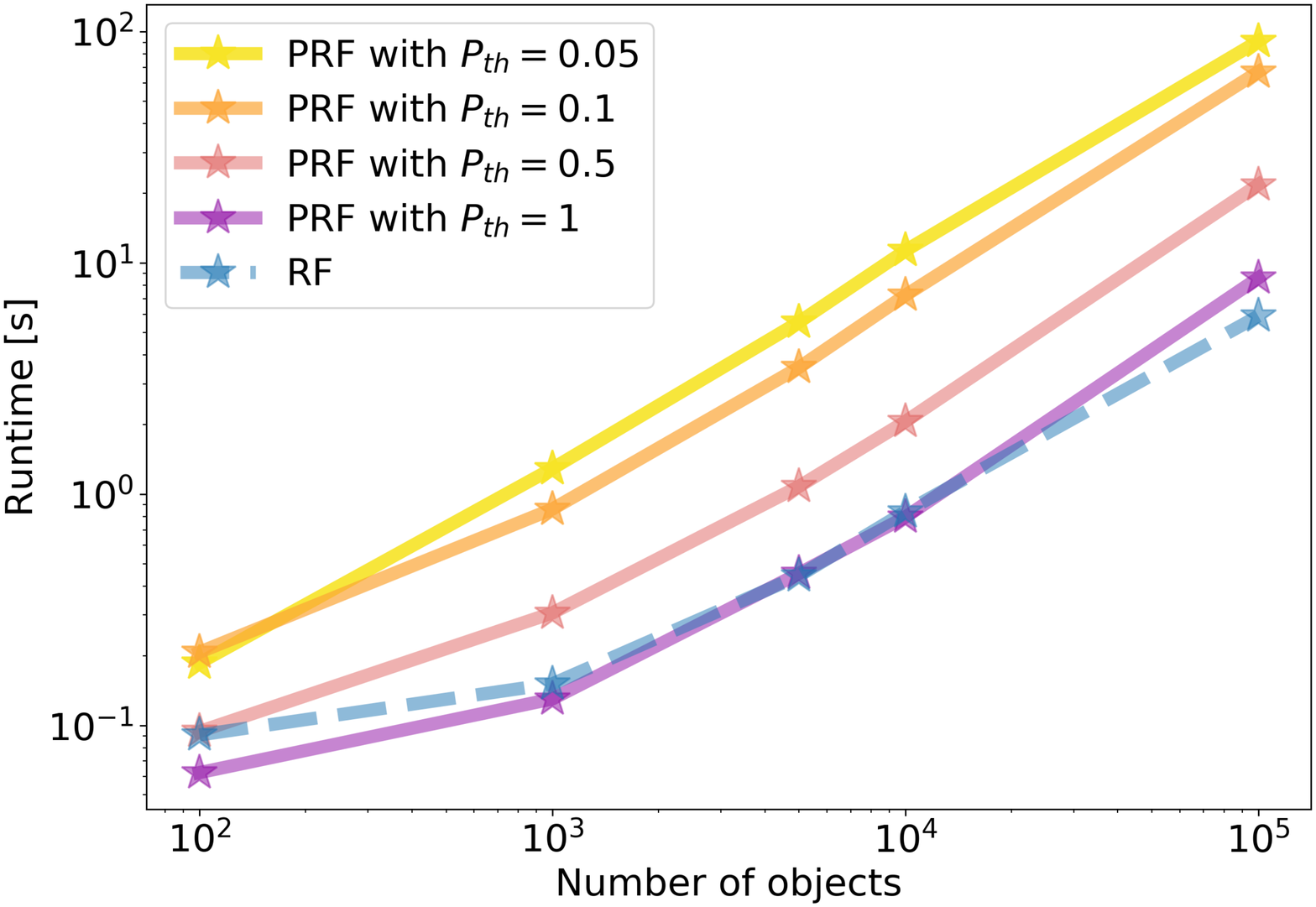}
  \caption{The runtime of the PRF algorithm as a function of the number of objects in the training and test sets, for different values of the $p_{th}$ parameter. The RF line (blue) refers to the {\fontfamily{cmtt}\selectfont  scikit-learn} implementation of RF. According to Figure \ref{fig:keep_proba}, the classification accuracy of the PRF converges to a constant value at $p_{th} = 5\%$.} 
  \label{fig:runtime}
  \end{center}
\end{figure}

\paragraph    {Runtime}
The runtime of the PRF algorithm is highly dependent on the PRF probability threshold parameter. Figure \ref{fig:runtime} shows the dependence of the runtime on the number of objects, for a regular RF, and PRF with different values of $p_{th}$. The number of objects in the training set is equal to number of objects in the test set, and this number is the one shown in the figure. Compared with Figure \ref{fig:keep_proba}, it could be seen that in order to achieve the maximal improvement in the classification accuracy, an increase in runtime of about an order of magnitude is needed. We note that as the RF is fast, such runtime should still be practical in most applications. Decreasing $p_{th}$ results in smaller increase in runtime but a more moderate improvement in accuracy (see Figure \ref{fig:keep_proba}). Investigating ways to further reduce the runtime is a prospect for future work.

\paragraph    {Best split search}\label{app:find_best_split}

\begin{figure}
  \begin{center}
  \includegraphics[width=0.5\columnwidth]{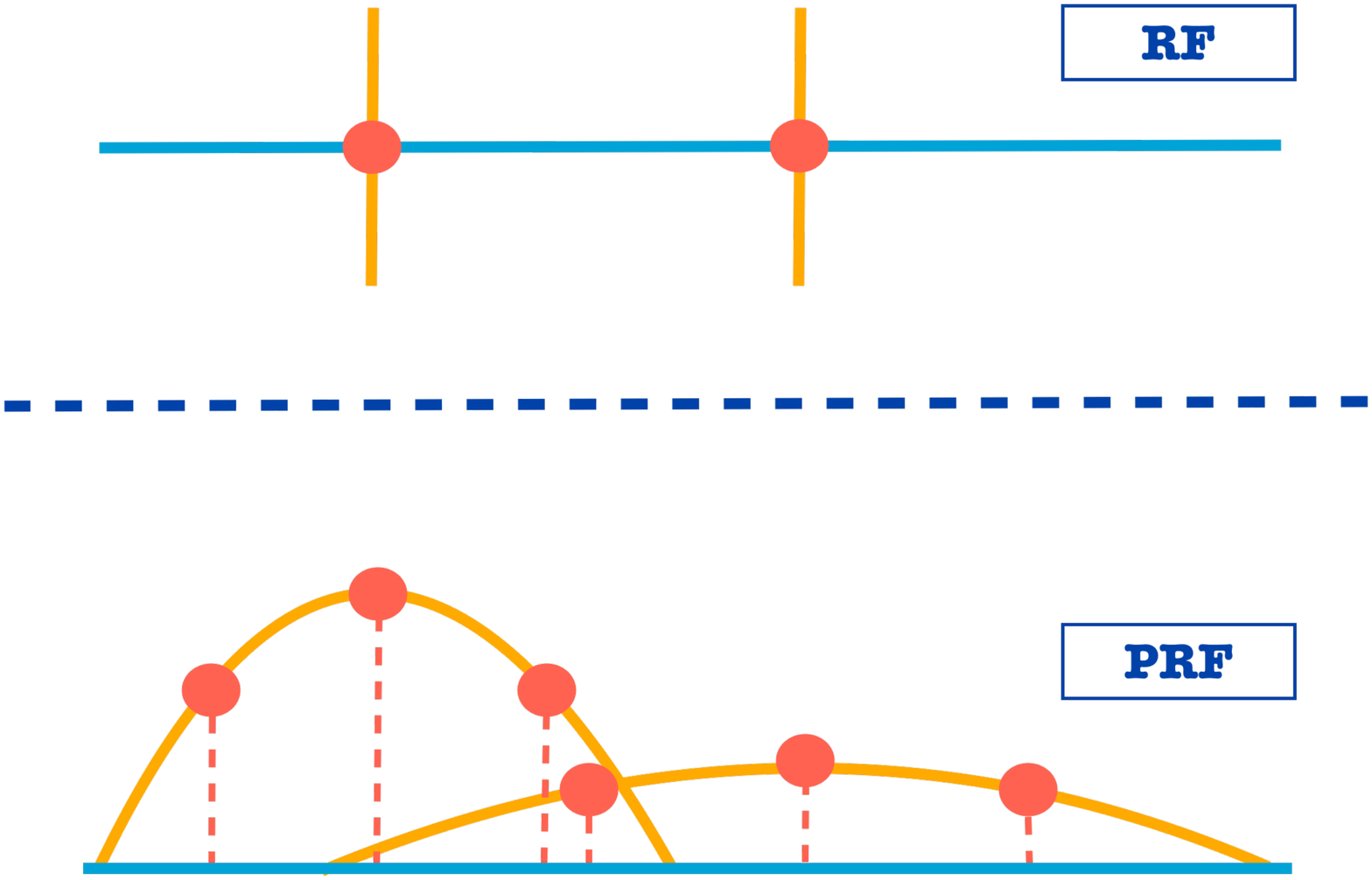}
  \caption{Illustration of the best split search grid in the  RF and PRF algorithms. The orange line represent the measurement (RF) and measurement PDF (PRF), while the red dots represent the grid. In the RF, the grid is constructed from the measurements values themselves. In the PRF, in order to use the information in the uncertainties,  points from defined locations on the PDFs of the measurements (e.g, $\pm \sigma$ from the center, $\sigma$ being the measurement uncertainty)  are added to the grid.}\label{fig:grid}
  \end{center}
\end{figure}

We start by describing the grid on which we search for the best split. In the original RF, the best split is defined as the threshold that results in the minimal combined impurity of the two groups (see equation \ref{eq:split_gini}). Thus, the grid on which the best split is searched is composed of the feature values themselves. That is, each feature value represents a single grid point, and the RF iterates over the grid points and finds the optimal threshold. This is illustrated in the top panel of Figure \ref{fig:grid}. In the PRF, since the features are PDFs, we choose to construct a grid that consists of the PDF's mean (i.e., the supplied feature value), and the $\pm 1\sigma$, $\pm 2\sigma$, and $\pm 3\sigma$ values around the mean (given by the feature uncertainty). This is illustrated in the bottom panel of Figure \ref{fig:grid}. We experimented with finer grids, and did not find an improvement in the classification accuracy. 

\begin{figure}
  \begin{center}
  \includegraphics[width=0.5\columnwidth]{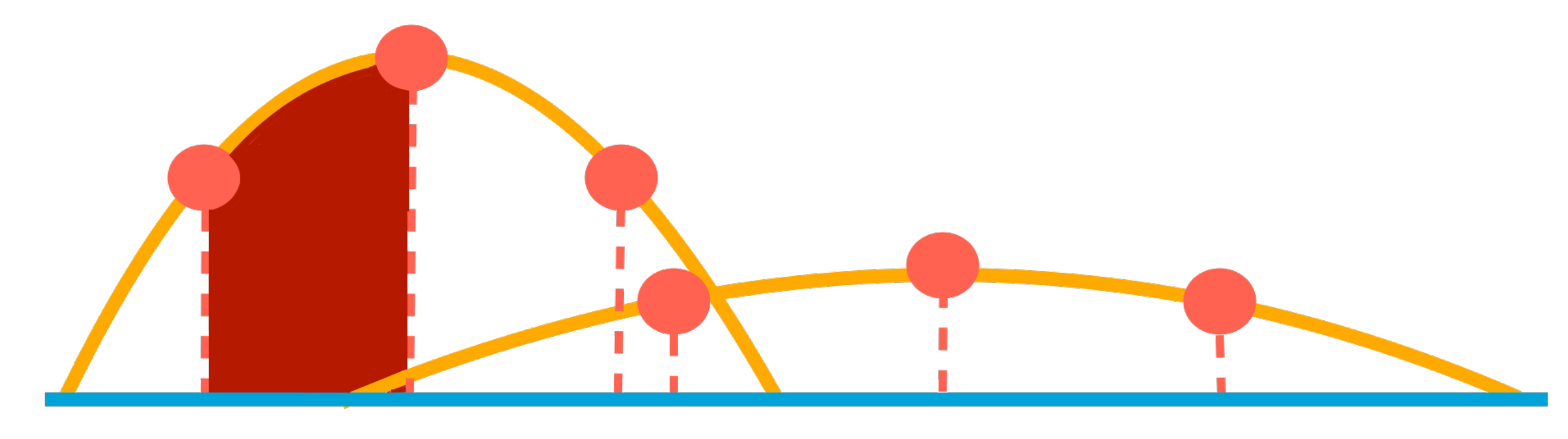}
  \caption{An illustration of the best split search grid of the PRF as described in Figure \ref{fig:grid}. Here a PDF 'chunk' is highlighted in dark red. In our implementation of the best split search we calculate the class probabilities for each point on the grid (red dots), for both the right and left sides of the point. Instead of calculating these from scratch for each point, we  traverse the grid sequentially from one side to the other, and for each new point  add the relevant PDF chunk to one side of the split and subtract it from the other.} 
  \label{fig:grid_v2}
  \end{center}
\end{figure}

The PRF iterates over the defined grid points, and the best split is the threshold that minimizes the combined impurity according to equation \ref{eq:13}. In order to compute the combined impurity, we must first sum over the PDFs of all the objects, where each PDF is weighted by the probability of each object to reach the given node (see equations \ref{eq:6}--\ref{eq:13}). One can, in principle, recompute the weighted sum of the PDFs at each iteration, by summing the commutative distributions of all the objects. However, we find that summing over all objects in each iteration takes a significant amount of computational time. Instead, we use a different approach. We start from the leftmost point on the grid and calculate the sum of the probabilities for this case (i.e., all the objects are on the left side of the split). When the threshold is shifted to the next grid point (one point to the right), instead of recomputing the summed probabilities, we only consider the part of the PDF that moved from the left group to the right group. We subtract this probability from the summed PDFs on the left side, and add it to the summed PDFs on the right side. The part of the PDF that moved from the left to the right is the integrated PDF between two grid points associated with the given object, as illustrated in Figure \ref{fig:grid_v2}. We note that this implementation is similar to the RF implementation in {\fontfamily{cmtt}\selectfont scikit-learn} version 0.19.2.

\bibliography{prf}



\end{document}